\documentclass[12pt]{article}
\usepackage{fullpage, natbib, authblk, hyperref}
\usepackage{amsfonts, amsmath, amssymb, amsthm, amsbsy, bm}
\usepackage{rotating, caption, subcaption, wrapfig, multirow}
\usepackage{graphicx}
\usepackage{floatrow}
\newfloatcommand{capbtabbox}{table}[][\FBwidth]
\hypersetup{
  pdftitle = {{B}ayesian estimation of in-game win probability for college basketball},
  pdfauthor = {Jason Maddox, Ryan Sides, and Jane L. Harvill},
  pdfsubject = {Bayesian analysis, estimating probability},
}

\begin{document}

\title{Bayesian estimation of in-game home team win probability for college basketball}

\author[1]{Jason T.~Maddox}
\author[2]{Ryan Sides}
\author[1]{Jane L.~Harvill} 
\affil[1]{\protect\raggedright 
Baylor University, Statistical Science, Waco, Texas, U.S., e-mail: jason\_maddox@baylor.edu}
\affil[2]{\protect\raggedright
Texas Woman's University, Mathematics and Computer Science, Denton, TX, U.S., e-mail: rsides@twu.edu}
	
	
\maketitle

\abstract{Two new Bayesian methods for estimating and predicting in-game home team win probabilities are proposed.  The first method has a prior that adjusts as a function of lead differential and time elapsed.  The second is an adjusted version of the first, where the adjustment is a linear combination of the Bayesian estimator with a time-weighted pre-game win probability.  The proposed methods are compared to existing methods, showing the new methods perform better for both estimation and prediction.  The utility is illustrated via an application to the 2016 NCAA Division 1 Championship game.}


\section{Introduction}
Sports analytics is not a new science.  Beginning with the work of pioneers like Bill James,  the work spans more than 30 years and a large range of difficulty.  Since the early 2000s, research in statistical methods for sports analytics has risen dramatically.  The review articles of~\citet{Kubatko_etal_2007}, \citet{fernandez_2019}, and~\citet{terner_franks_2020} provide a fairly comprehensive review for sports analytics for a wide variety of sports, including basketball.

Generally speaking, models for predicting the outcome of a sporting event can be classified into two systems: (1) pre-game prediction or (2) in-game, or in-play, prediction.  Pre-game prediction involves determining the outcome of a game before play begins.  In contrast, in-game prediction attempts to use the progress during a game to determine win probabilities that vary as a function of in-game variables, for example, elapsed game time, or score difference.  The focus of this paper is in-game prediction, and more specifically, during the course of the game estimating the probability the home team wins, or the ``in-game win probability.''  

Efforts to accurately estimate in-game win probability have long been a pastime and many different methods for doing so can be found in the literature.  \citet{cooper_deneve_mosteller_1992} collected and analyzed data from 200 basketball games, 100 baseball games, 100 hockey games, and 100 football games to investigate when, during the course of a game for each of the sports, it is most likely to know the final outcome of the game.  With respect to basketball they concluded, ``late game leaders in basketball go on to win about four in five times,'' and that percentage is ``no different in football or hockey.''  They also found that ``home teams in basketball were more than three times as likely as visiting teams to make a fourth-quarter comeback,'' and that the home winning percentage for basketball (64.1\%) was the strongest evidence of a home-team advantage across major American sports.

One of the earlier papers on predicting in-game win probability for baseball is attributed to \citet{lindsey_1963} who used the maximum likelihood estimator to determine the probability the home team wins given the inning the game is in and the home team's lead.  He looked into how many times a team had won in prior games when in the same position as the current game was in. That information was then used to determine that team's empirical winning percentage.  Up until his work, baseball decisions were based on what would maximize the scoring output for an inning.  \citeauthor{lindsey_1963}'s groundbreaking research instead focused on determining how each decision would affect the probability that a team wins instead of only the change in expected score.  A more recent development in estimating in-game win probability is  \citet{lock_nettleton_2014} who use random forests that combine pre-play variables to estimate win probability before each play of an NFL game.  Additionally, \citet{ryall-2011} used play-by-play data with pre-game Elo rankings to develop a model for Australian Rules football.  The concept of using pre-game power rankings is one that will be adopted here.  
 
In-game basketball analytics began to surface when \citet{westfall_1990} developed a graphical summary of the scoring activity for a basketball game that is a real-time plot of the score difference versus the elapsed time.  The features of the graph provided easy access to largest leads, lead changes, come-from-behind activity, and other interesting game features.   Over time, models for forecasting in-game win probability have become more complex.  Some are built on expert predictions, some on betting paradigms, and others on within-game metrics.  \citet{shirley_2007} modeled a basketball game using a Markov model with three states and used that model for estimating in-game win probability.  \citet{strumbelj_vracar_2012} improved upon the model of \citeauthor{shirley_2007} by taking into consideration the strengths of the two teams and estimated transition probabilities using performance statistics.  \citet{vracar-etal-2016} extended the state description to capture other facets beyond in-game states so that the transition probabilities become conditional on a broader game context. \citet{bashuk_2012} proposed using cumulative win probabilities over the duration of a game to measure team performance.  Using five years of game play, he generated a Win Probability Index for NCAA basketball.  Using that he created an open system to measure the impact, in terms of win probability added, of each play. 

More recently, \citet{chen-fan-2018} developed a method for predicting game outcomes using a functional data approach.  \citet{shi_song_2019} develop a discrete-time, finite-state Markov model for the progress of basketball scores, and use it to conditionally predict the probability the home team wins or loses by a certain amount.  \citet{song_shi_2020} present an in-play prediction model based on the gamma process.  They apply a Bayesian dynamic forecasting procedure that can be used to predict the final score and total points.  Finally,~\citet{song-gao-shi-2020} modify the gamma process by employing betting lines, letting the expectation of the final points total equal the pre-game betting line. 

While the techniques outlined above have various pros and cons, the methods for estimating the in-game win probability proposed here are aimed at improving the approaches of \citet{stern_1994} and \citet{deshpande_jensen_2016}. \citet{stern_1994} modeled the difference between the home and visiting teams' scores as a Brownian motion process with drift equal to the points in favor of the home team.  This model, which is equivalent to a probit regression model, results in a relationship between the home team's lead and the probability of victory for the home team.  His approach is one of the earliest to provide a mechanism for allowing time to be continuous, and to use that continuity in modeling the probabilities. \citet{deshpande_jensen_2016} extended the work of \citeauthor{stern_1994} by applying a Bayesian framework to the probit model.

The remainder of the paper is organized as follows.  Section~\ref{sec:models} contains a more thorough overview of \citet{stern_1994} and \citet{deshpande_jensen_2016}, two methods for in-game probability designed specifically for NBA games.  In Sections~\ref{sec:dynamicprior} and~\ref{sec:adjustment}, a modified enhanced Bayesian approach is proposed that not only improves in-game predictions compared to existing methods, but also is suitable for application to NCAA basketball.  Section~\ref{sec:sim} presents a description of the data used in the study.  Following that description is the result of estimating the in-game home team win probability to over 30,000 NCAA basketball games, and then using the estimated model to predict the outcome of over 10,000 NCAA games.  To illustrate utility in Section~\ref{sec:application}, the models are applied to the 2016 Division 1 NCAA Tournament Championship game between the University of North Carolina and Villanova University. Finally, Section~\ref{sec:conclusion} contains a summary and concluding remarks.

\section{Estimating In-Game Win Probability}
\label{sec:models}
For a specific game, consider the random process that is the home team's lead at time $t = 0, 1, \ldots, 2399$, where $t$ is the game time elapsed in seconds.  At a specific time $t$ and for a specific home team lead $\ell$, let $p_{t,\ell}$ denote the in-game probability that the home team will win the game at the end of regulation.  At the beginning of the game, $t = \ell = 0$, the estimator of $p_{0,0}$ is dependent on the method used for estimating in-game win probability.

When considering multiple games $i = 1, 2, \ldots, M$, let $Y_i = 1$ if the home team wins game $i$ and 0 otherwise.  Consider $p_{t,\ell}$ as a continuous function of $t$ and $\ell$.  A classic approach to estimating $p_{t,\ell}$ in any $(t, \ell)$ cell is the maximum likelihood estimator.  Specifically, define $N_{t,\ell}$ as the number of games in which the home team leads by $\ell$ points after $t$ seconds, and define $n_{t,\ell} = \sum_{i=1}^{N_{t,\ell}} Y_i$, the number of games in the $(t, \ell)$ cell that the home team wins in regulation.  Then on each $(t, \ell)$ cell, the random variable $n_{t,\ell}$ has a binomial($N_{t,\ell}, p_{t,\ell})$ distribution. Within each cell, the maximum likelihood estimator of $p_{t, \ell}$ is $\bar p_{t,\ell} = n_{t,\ell}/N_{t,\ell}$.  At $(0,0),~\bar p_{0,0} = n_{0,0}/N_{0,0}$.  

Let $X_t$ represent the home team lead after $t$ seconds.  Another approach to estimating $p_{t,\ell}$ is found in \citet{stern_1994}.  \citeauthor{stern_1994} estimates in-game home team win probability via a Brownian motion process with drift $\mu$ points per second home team lead and finite variance $\sigma^2$; that is,
\begin{equation}
    \label{eq:brownianmotion}
    \tilde p_{t^*,\ell} = P(X_1 > 0\,|\,X_{t^*} = \ell) = \Phi\left(\frac{\ell + (1 - t^*)\mu}{\sqrt{(1 - t^*)\sigma^2}}\right),
\end{equation}
where $t^* \in [0,1)$ represents re-scaled time $t$ and $\Phi(\cdot)$ denotes the standard normal cumulative distribution function.  Although $X_t$ is discrete, the model in equation~\eqref{eq:brownianmotion} treats $X_t$ as a continuous random variable.  \citeauthor{stern_1994} suggests a continuity correction factor be applied $\ell$, although he also noted the continuity correction factor results in little improvement in the model's performance.  He provides empirical evidence that the Brownian motion model provides a good fit to the score differences when applied across multiple games.  \citeauthor{stern_1994} noted the model~\eqref{eq:brownianmotion} can be interpreted as a probit regression model relating the game outcome to the transformed variables $\ell/\sqrt{1 - t^*}$ and $\sqrt{1 - t^*}$, and with coefficients $\sigma^{-1}$ and $\mu/\sigma$.  
At $t^* = 0$ and  $\ell = 0,~\tilde p_{0,0} = \Phi(\mu/\sigma)$, signifying that $\mu/\sigma$ indicates the magnitude of the home field advantage.  Specifying the home team advantage can be accomplished several ways.  \citet{stern_1994} suggests letting $\mu$ be the home field advantage in the particular sport, and for basketball, $\mu = 5$ or 6 points.  They also note that the home team wins in approximately 55\% to 65\% of games, and so $\sigma$ can be chosen so that values of $\mu/\sigma$ are in the 0.12 to 0.39 range. 

Some $(t, \ell)$ cells may have small values of $N_{t, \ell}$, which have the potential to result in estimators of $p_{t,\ell}$ with very large standard errors.  To address this issue, windows can be defined and centered on $(t,\ell)$ in such a way that the in-game win probability remains relatively constant across the window.  In basketball, since no offensive possession can result in more than four points the window with respect to $\ell$ can be reasonably defined as $[\ell - 2, \ell + 2]$.  Moreover, since most offensive possessions last at least six seconds the width of the time window is taken to be six.  The same notation will be adopted for any $[t - 3, t + 3] \times [\ell - 2, \ell + 2]$ window; that is, $N_{t,\ell}$ is the number of games in the window in which the home team has led by any value in $[\ell - 2, \ell + 2]$ points after any time in $[t - 3, t + 3]$ seconds and $n_{t,\ell} = \sum_{i=1}^{N_{t,\ell}} Y_i$, distributed as a binomial($N_{t,\ell}, p_{t,\ell})$ random variable.

Bayesian methods provide a third approach to estimating $p_{t,\ell}$.  It is well-known that the beta family is a conjugate prior for estimating $p_{t, \ell}$.  Let $\alpha_{t,\ell} > 0$ and $\beta_{t,\ell} > 0$ be the shape parameters of the beta prior on $p_{t,\ell}$.  Then for each window within the $(t, \ell)$ plane, the Bayes estimator of $p_{t,\ell}$ is the mean of the posterior; specifically
\begin{equation}
    \label{eq:bayes}
    \hat p_{t,\ell} = \frac{n_{t,\ell} + \alpha_{t,\ell}}{N_{t,\ell} + \alpha_{t,\ell} + \beta_{t,\ell}}.
\end{equation}
For all $t = 0, 1, \ldots, M$, on all windows centered at $(t, \ell)$,  \citet{deshpande_jensen_2016} propose the beta prior
\begin{equation}
\label{eq:djprior}
    p_{t, \ell} \sim \left\{
    \begin{array}{ll}
    \text{beta}(0, 10), & {\text{for $\ell < -20$,}} \\
    \text{beta}(5, 5),  & {\text{for $-20 \leq \ell \leq 20$,}} \\
    \text{beta}(10, 0), & {\text{for $\ell > 20$.}}
    \end{array}\right.
\end{equation}
The choice of prior depends only on the home team lead $\ell$, and does not take into account the time remaining in the game $t$.

\subsection{Dynamic Prior for In-Game Home Team Win Probability}
\label{sec:dynamicprior}
\begin{figure}[!ht]
\caption{Illustration of dynamic beta prior.}
\includegraphics[width=0.8\textwidth]{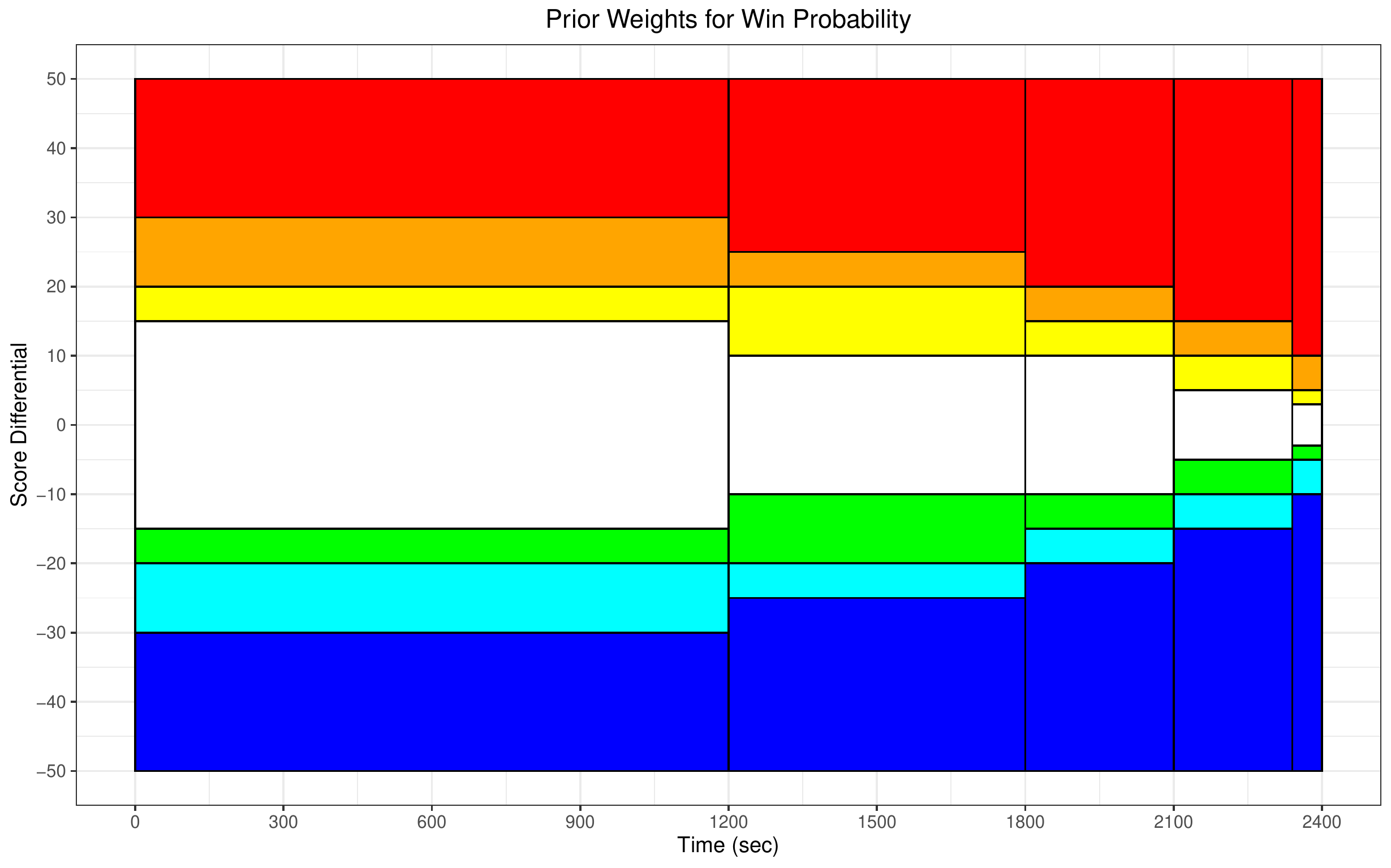}
\label{fig:dynamicprior}
\end{figure}

The scale parameters in~\eqref{eq:djprior} rely only on the home team lead $\ell$, not taking into account the interaction between $\ell$ and time remaining. Also notable is that for $|\ell| > 20$, the prior parameters yields an improper prior.  In these cases, the prior may overwhelm the information in the data.  To model the interaction between the home team lead and the time elapsed, we propose choosing the scale parameters for the beta prior dynamically, as illustrated in Figure~\ref{fig:dynamicprior} and specified in Table~\ref{tab:dynamicprior}.
\begin{table}[!ht]
\caption{Specification of dynamic beta prior.}
\begin{tabular}{rl}
Color      & Prior \\ \hline
Red        & beta(19,1) \\
Orange     & beta(9,1)  \\
Yellow     & beta(4,1)  \\
White      & beta(1, 1) \\
Green      & beta(1, 4) \\
Light blue & beta(1, 9) \\
Blue       & beta(1, 19) \\ 
\end{tabular}
\label{tab:dynamicprior}
\end{table}

The time intervals illustrated in Figure~\ref{fig:dynamicprior} are $[0, 1200), [1200, 1800), [1800, 2100), [2100, 2340)$, and $[2340, 2400]$.  Observed home-team differential intervals are time dependent.  For the small number of games when the $|\ell| > 50$, the beta(19, 1) or beta(1, 19) prior is applied accordingly.  In the rare instance that a specific $(t, \ell)$ window had no observed data, for a positive differential, the largest posterior probability of a home team win is used; for a negative differential, the smallest posterior probability of a home team win is used.  This choice of posterior estimates is justified since windows with empty cells are those windows with large score differentials relative to the elapsed time.  Moreover, that the combination hasn't occurred in over 30,000 games.  Therefore, the choice of the largest or smallest estimate is reasonable.  Using the dynamic prior, for any $(t, \ell)$ in the plane, the dynamic Bayesian estimator of $p_{t,\ell}$ is given as in equation~\eqref{eq:bayes}.

The newly proposed choice of prior is proper for all $(t, \ell)$ combinations and is much less likely to overwhelm the information in the data.  More importantly, the dynamic prior models the in-game home team win probability as a combination of $(t, \ell)$.  As with the prior in~\eqref{eq:djprior}, the size of the score differential $|\ell|$ has an affect on the estimated in-game win probability.  However, the newly proposed dynamic prior also models the estimated in-game win probability as a function of the game time elapsed.  In particular, early in the game (for small $t$), even for a relatively large score differential, the dynamic prior allows for a larger probability of a comeback than late in the game, for the same differential.  

\subsection{Adjustment of Bayesian Estimator}
\label{sec:adjustment} 
For any game, the skill of the two teams playing has a definitive impact on how likely it is a team will win.  If two teams are evenly matched, then accurately predicting which team will win is more difficult than if one team is much more skilled than the other.  Accordingly, team skill should also be taken into account when estimating the in-game win probability.  None of the previously described estimators incorporate such a measure.  In the following, we propose an adjusted Bayesian estimator for in-game win probability, where the adjustment incorporates the skills of the teams in the game.

To determine the adjustment, or the pre-game point spread, the team ratings and home team advantage are used. The point differential is commonly modeled using a normal distribution with a standard deviation of 10 points; see for example,~\citet{adams_2019}. The mean for each game is computed using the difference in the two teams' ratings with a 3.5 point advantage added for the home team.  Using the normal quantile function, the pre-game point spread can be converted to a pre-game winning probability, $\hat p_p$ say, for each team.  For the rare case when a non Division 1 team appeared in the database, that team is given a rating slightly lower than the lowest ranked Division 1 team.  While this may not be accurate for each of these occurrences, no power rankings on Division 2 and lower teams are available.  During the game at each time $t$, the probability the home team wins $\hat p^*_{t, \ell}$ is the weighted average of the pre-game win probability and the current unadjusted in-game probability, $\hat p_{t, \ell}$ found using the dynamic Bayesian model described in Section~\ref{sec:dynamicprior}; that is,
\begin{equation}
    \label{eq:best}
    \hat p^*_{t, \ell} = \left(\frac{2400 - t}{2400}\right)\hat p_p + \left(\frac{t}{2400}\right)\hat p_{t, \ell}
\end{equation}
The adjustment has the following effect. In a tight contest, the in-game win probability is shifted such that the higher ranked team is predicted to be more likely to win.  When the higher ranked team is winning, the probability the higher ranked time will win is larger than a lower ranked team that is winning at the same time by the same margin.

\section{Comparison of Methods for Estimation and Prediction}
\label{sec:sim}
In the following, the performances of the MLE, the probit model estimator, the Bayesian estimator with prior given in~\eqref{eq:djprior}, and the Bayesian estimator with dynamic prior in Table~\ref{tab:dynamicprior} and Figure~\ref{fig:dynamicprior} are compared from both the modeling and prediction perspectives.  The games that are used in conducting the study are from the 2012-2013 seasons up until the 2019-2020 season, which was cut short due to COVID.  In Subsection~\ref{sec:data}, the data is more thoroughly described, and the data collection process and its challenges are discussed.  In Section~\ref{sec:estimation} the models are used to compute estimates of in-game home team win probability using 30,789 games beginning with the 2012-2013 season and ending with the 2017-2018 season.  In Section~\ref{sec:prediction}, the model estimates are used to predict the outcome of the 10,853 games from the 2018-2019 season and the 2019-2020 season, which was cut short due to COVID.  The performances of the models are compared through Brier's score and misclassification rates.  

\subsection{NCAA Data, Collection and Challenges}
\label{sec:data}
Play-by-play data from ESPN was scraped using \emph{R} and the package {\tt rvest}.  Since play-by-play data is not readily available on ESPN's site for college basketball games prior to 2012, the data collected begins with the 2012-2013 season and continues through the completion of the 2019-2020 season, when the season was cut short due to the COVID-19 pandemic.

Due to inconsistencies in data collection and formatting, the data used does not contain play-by-play data for every game for all teams.  From each conference, a randomly selected team was chosen for inclusion in Table~\ref{tab:data}, which shows the percentage of all games played for each season in the data. In general, smaller schools have a lower percentage of games with available data than larger schools. However, the lack of information did not appear systematic, and is not seen as problematic, since the data does include play-by-play information for many other similar teams.

\begin{table}[!ht]
\caption{Percentage of games played for which data was collected by ESPN for selected programs from each conference.}
\tiny
\begin{tabular}{rrccccccccc} 
Team            & Conference      & 12-13 & 13-14 & 14-15 & 15-16 & 16-17 & 17-18 & 18-19 & 19-20 & Total  \\ \hline
Cent Arkansas   & ASUN            & 76.7\%  & 93.1\%  & 79.3\%  & 100.0\% & 100.0\% & 91.4\%  & 93.9\%  & 90.3\%  & 90.6\% \\
UMass Lowell    & America East    & NA      & 89.3\%  & 89.7\%  & 96.6\%  & 93.5\%  & 93.3\%  & 96.9\%  & 90.6\%  & 92.8\% \\
UCF             & American        & 90.0\%  & 93.5\%  & 93.3\%  & 96.7\%  & 97.2\%  & 96.9\%  & 87.9\%  & 96.7\%  & 94.0\% \\
George Mason    & Atlantic 10     & 78.9\%  & 87.1\%  & 87.1\%  & 100.0\% & 88.2\%  & 97.0\%  & 93.9\%  & 96.9\%  & 91.1\% \\
NC State        & ACC             & 17.1\%  & 94.4\%  & 97.2\%  & 97.0\%  & 100.0\% & 97.0\%  & 100.0\% & 100.0\% & 87.8\% \\
Texas Tech      & Big 12          & 19.4\%  & 90.6\%  & 96.9\%  & 93.8\%  & 96.9\%  & 97.3\%  & 97.4\%  & 93.5\%  & 85.7\% \\
Xavier          & Big East        & 67.7\%  & 91.2\%  & 83.8\%  & 100.0\% & 89.5\%  & 97.1\%  & 97.1\%  & 96.9\%  & 90.4\% \\
N Arizona       & Big Sky         & 87.5\%  & 90.6\%  & 92.1\%  & 86.7\%  & 100.0\% & 93.8\%  & 96.8\%  & 86.7\%  & 91.8\% \\
Gardner-Webb    & Big South       & 85.3\%  & 90.9\%  & 94.3\%  & 87.9\%  & 93.9\%  & 96.9\%  & 85.7\%  & 90.9\%  & 90.7\% \\
Ohio State      & Big Ten         & 10.8\%  & 85.7\%  & 85.7\%  & 100.0\% & 100.0\% & 97.1\%  & 97.1\%  & 100.0\% & 84.6\% \\
UC Irvine       & Big West        & 75.7\%  & 91.4\%  & 85.3\%  & 86.8\%  & 91.7\%  & 91.4\%  & 89.2\%  & 100.0\% & 88.9\% \\
Delaware        & Colonial        & 60.6\%  & 94.3\%  & 90.0\%  & 86.7\%  & 100.0\% & 100.0\% & 90.9\%  & 90.9\%  & 89.2\% \\
UNC Charlotte   & Conference USA  & 75.0\%  & 90.3\%  & 90.6\%  & 93.9\%  & 96.7\%  & 93.1\%  & 93.1\%  & 100.0\% & 91.6\% \\
Oakland         & Horizon         & 81.8\%  & 87.9\%  & 90.9\%  & 94.3\%  & 97.1\%  & 97.0\%  & 100.0\% & 84.8\%  & 91.7\% \\
Harvard         & Ivy             & 69.0\%  & 90.6\%  & 80.0\%  & 90.0\%  & 96.4\%  & 100.0\% & 93.5\%  & 89.7\%  & 88.7\% \\
Fairfield       & MAAC            & 74.3\%  & 87.5\%  & 96.8\%  & 97.0\%  & 96.8\%  & 93.9\%  & 93.5\%  & 93.8\%  & 91.7\% \\
Kent State      & Mid-American    & 80.0\%  & 78.1\%  & 91.4\%  & 93.8\%  & 91.7\%  & 97.1\%  & 100.0\% & 96.9\%  & 91.1\% \\
Morgan State    & MEAC            & 62.5\%  & 41.9\%  & 58.1\%  & 41.9\%  & 63.3\%  & 81.3\%  & 83.3\%  & 71.0\%  & 62.9\% \\
Drake           & Missouri Valley & 78.1\%  & 90.3\%  & 96.8\%  & 96.8\%  & 100.0\% & 97.1\%  & 97.1\%  & 97.1\%  & 94.1\% \\
San Diego State & Mountain West   & 44.1\%  & 91.7\%  & 94.4\%  & 92.1\%  & 90.9\%  & 93.9\%  & 94.1\%  & 100.0\% & 87.7\% \\
St Francis (PA) & Northeast       & 89.7\%  & 80.6\%  & 100.0\% & 96.7\%  & 97.1\%  & 96.8\%  & 90.9\%  & 84.4\%  & 92.0\% \\
SE Missouri St  & Ohio Valley     & 63.6\%  & 53.1\%  & 96.7\%  & 93.1\%  & 97.0\%  & 93.5\%  & 87.1\%  & 100.0\% & 85.5\% \\
Colorado        & Pac-12          & 36.4\%  & 94.3\%  & 90.9\%  & 94.1\%  & 94.1\%  & 93.8\%  & 97.2\%  & 93.8\%  & 86.8\% \\
Bucknell        & Patriot League  & 79.4\%  & 90.0\%  & 88.2\%  & 93.5\%  & 100.0\% & 97.1\%  & 97.0\%  & 97.1\%  & 92.8\% \\
Missouri        & SEC             & 32.4\%  & 94.3\%  & 96.9\%  & 100.0\% & 93.8\%  & 97.0\%  & 96.9\%  & 100.0\% & 88.9\% \\
Furman          & Southern        & 90.3\%  & 83.3\%  & 93.9\%  & 94.3\%  & 91.4\%  & 93.9\%  & 90.9\%  & 100.0\% & 92.3\% \\
McNeese State   & Southland       & 54.8\%  & 87.1\%  & 87.1\%  & 93.1\%  & 100.0\% & 96.4\%  & 93.5\%  & 90.6\%  & 87.8\% \\
Jackson State   & SWAC            & 41.4\%  & 32.3\%  & 43.8\%  & 41.7\%  & 34.4\%  & 50.0\%  & 40.6\%  & 90.6\%  & 46.8\% \\
Omaha           & Summit League   & 80.6\%  & 46.9\%  & 69.0\%  & 90.6\%  & 96.9\%  & 93.5\%  & 90.6\%  & 93.8\%  & 82.7\% \\
South Alabama   & Sun Belt        & 83.3\%  & 90.3\%  & 97.0\%  & 87.9\%  & 96.9\%  & 96.9\%  & 97.1\%  & 96.8\%  & 93.3\% \\
BYU             & West Coast      & 58.3\%  & 94.3\%  & 94.3\%  & 97.3\%  & 94.1\%  & 100.0\% & 90.6\%  & 100.0\% & 91.1\% \\
Chicago State   & WAC             & 75.8\%  & 90.6\%  & 62.5\%  & 87.5\%  & 100.0\% & 96.9\%  & 84.4\%  & 89.7\%  & 85.9\% \\
\end{tabular}
\label{tab:data}
\end{table}
Part of the play-by-play data includes the arena where the game was played.  Teams are assigned multiple home arenas in situations where that is appropriate due to design, i.e., Villanova, or because of a one off, for example, TCU playing their home games in a different arena for the 2014-2015 season due to renovations. While ESPN is not globally consistent in the way information is stored, this effect on home-arena determination is rare.  It should also be noted that occasionally teams play in what could be considered a home game with respect to fan attendance and location.  An example of this is when the University of Kansas plays in Kansas City.  Despite these issues, the database of games is large enough that the overall accuracy of the data used in the analysis is unaffected.

Finally, to capture each team's power rating prior to each game, data was scraped from the website {\tt teamrankings.com}.\footnote{{\tt teamrankings.com} lacks data for December 3, 2012.  For games played on that date, the team ratings from the previous day were used.}  This site was chosen over other systems that may outperform {\tt teamrankings.com} because {\tt teamrankings.com} has historical daily power ratings available and the other sites do not.  An alternative to using these ratings might be the betting spread from Las Vegas betting lines.

\subsection{Estimating Home Team In-Game Win Probability}
\label{sec:estimation}
\begin{figure}[!ht]
  \caption{Plots illustrating estimated in-game home team win probabilities based on maximum likelihood (top left), probit model (top right), traditional Bayes (bottom left), and Bayes with dynamic prior (bottom right).}
  \begin{subfigure}[t]{0.475\textwidth}
    \caption{Maximum likelihood estimates.} 
    \includegraphics[width=\textwidth]{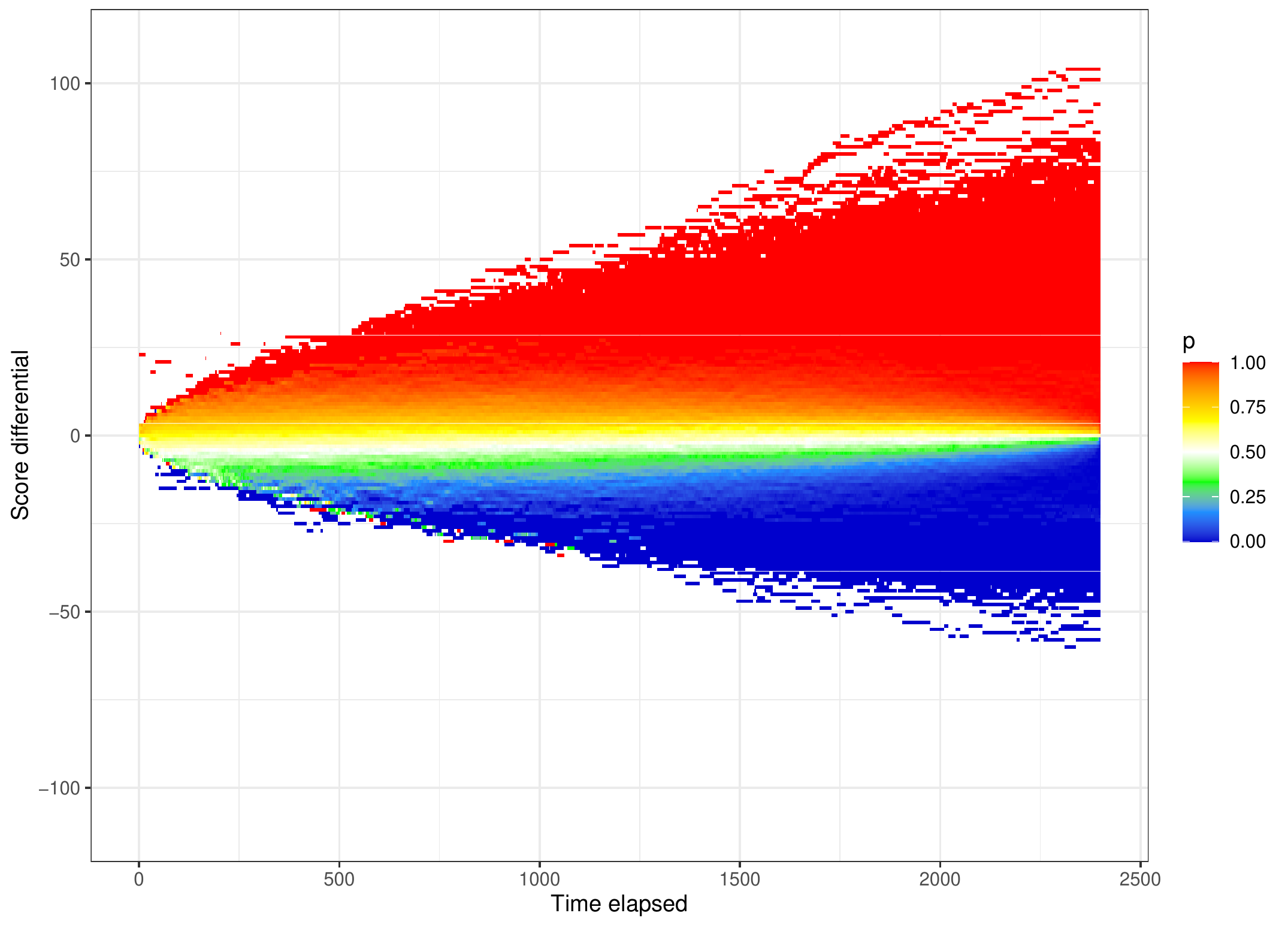}
    \label{fig:MLE}
  \end{subfigure}
  \hfill
  \begin{subfigure}[t]{0.475\textwidth}
    \caption{Probit model estimates.} 
    \includegraphics[width=\textwidth]{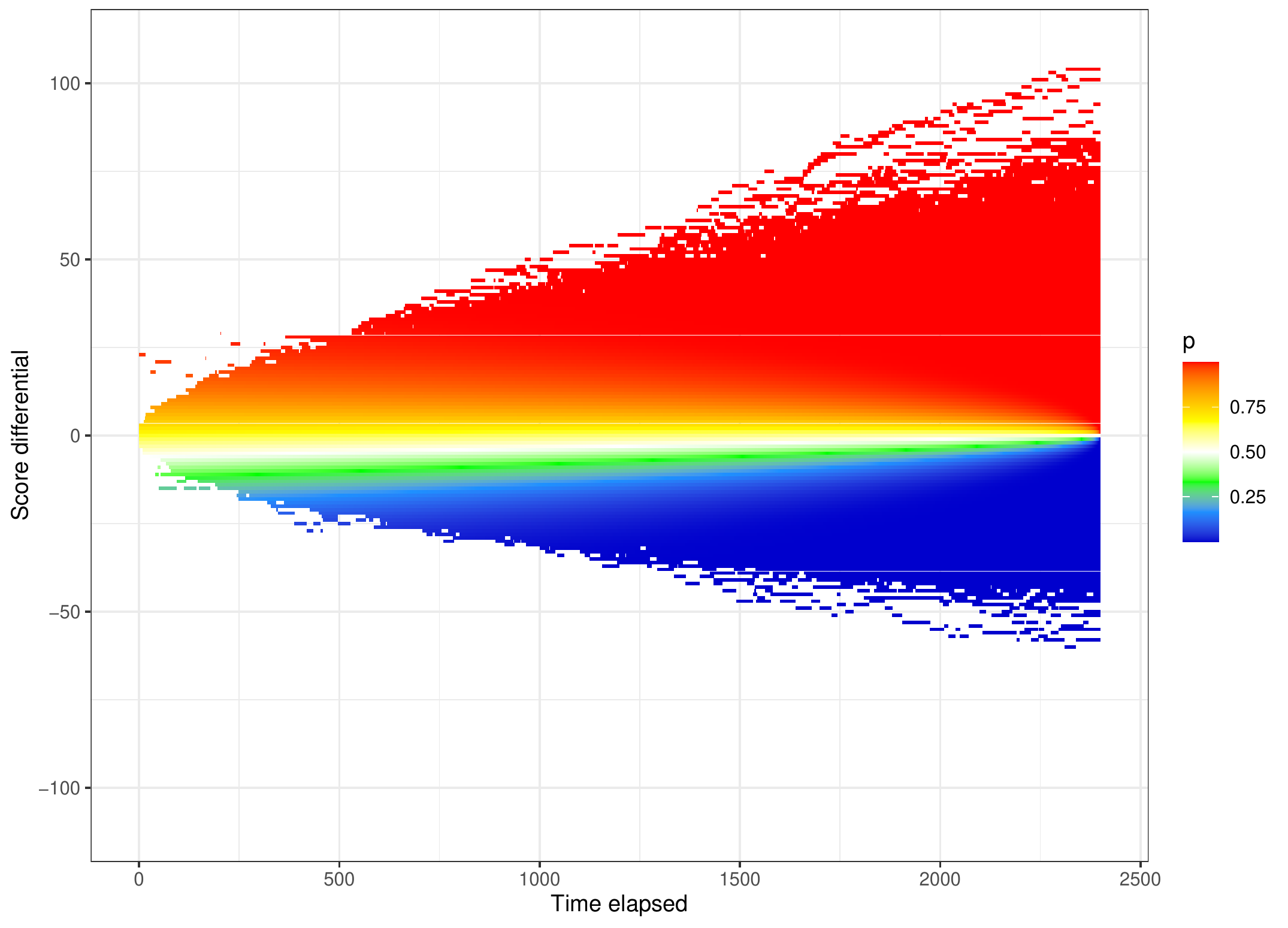}
    \label{fig:probit}
  \end{subfigure}
  
  \begin{subfigure}[t]{0.475\textwidth}
    \caption{Bayesian estimates.}
    \includegraphics[width=\textwidth]{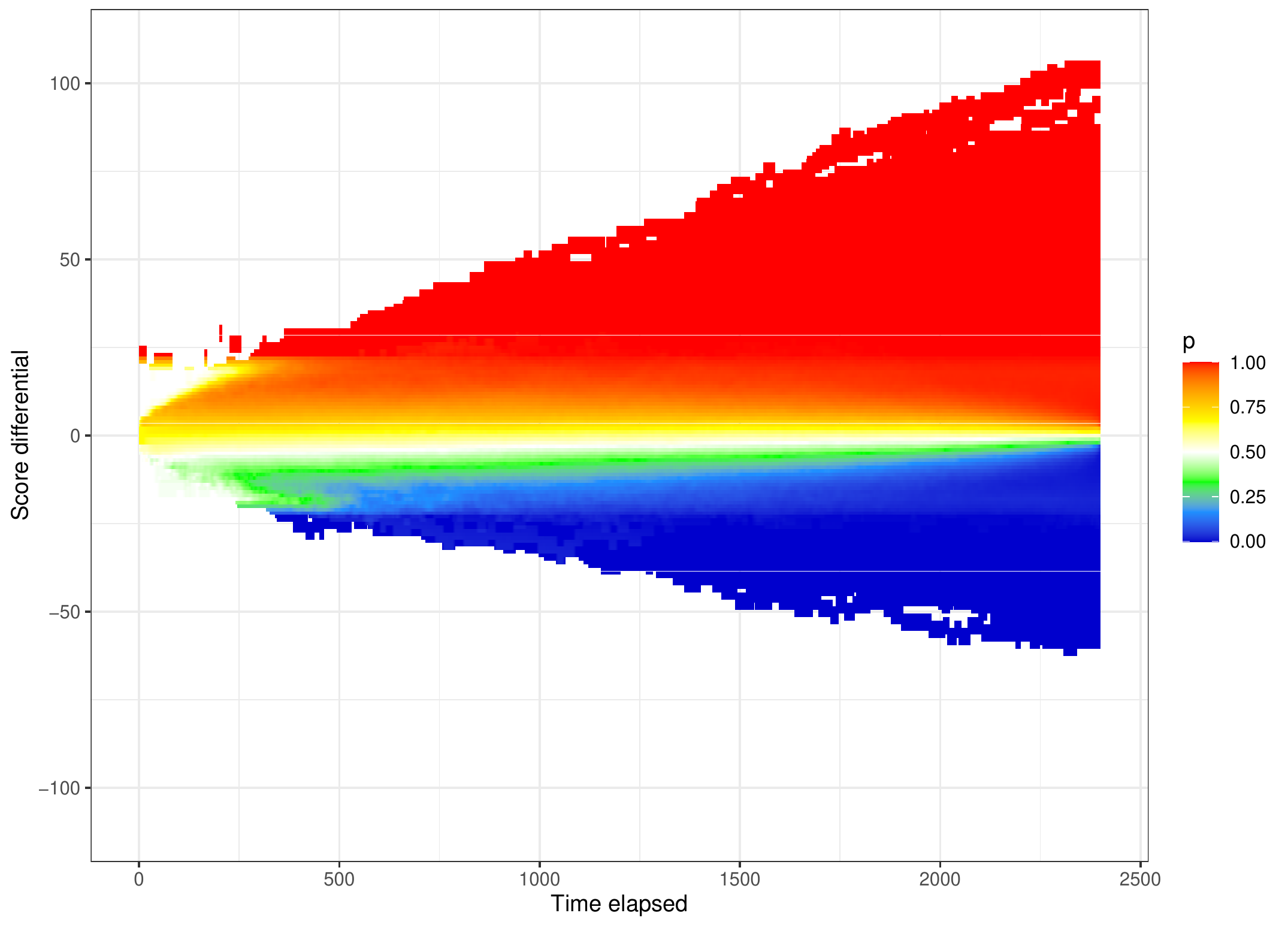}
    \label{fig:Bayes}
  \end{subfigure}
  \hfill
  \begin{subfigure}[t]{0.475\textwidth}
    \caption{Dynamic Bayesian estimates.}
    \includegraphics[width=\textwidth]{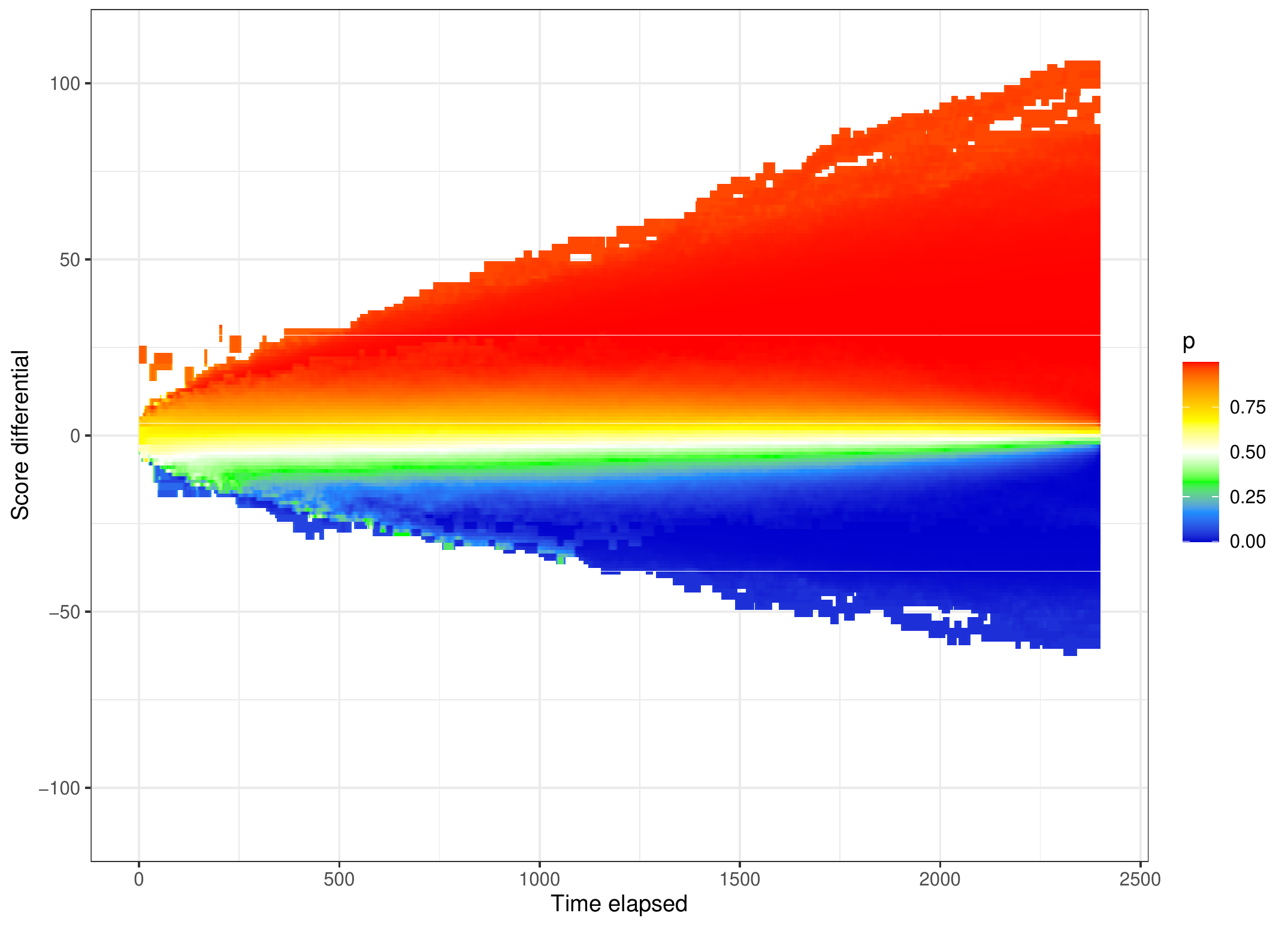} 
    \label{fig:dynamicBayes}
  \end{subfigure}
  \label{fig:probests}
\end{figure}

Estimates of in-game home team win probability from the MLE, probit model, Bayes model using the prior in~\eqref{eq:djprior}, and the dynamic Bayes model are given in Figures~\ref{fig:MLE} through~\ref{fig:dynamicBayes}.  For each $t = 0, 1, \ldots, 2399$ and $|\ell| = 0, 1, \ldots, 104$ cell, the estimates are illustrated by letting the color of any $(t, \ell)$ represent the estimated value of $p_{t,\ell}$.  The values $|\ell| = 104$ are taken from a game played on Dec.~30, 2013, when Southern University beat Champion Baptist College by a score of 116-12, the largest margin of victory for any game included in the data described in Section~\ref{sec:data}.  Historically, the largest margin of defeat in college men's basketball occurred on Jan.~12, 1992 when Troy State University, the home team, beat DeVry University of Atlanta by a score of 258-141.  For any cell, blue represents an estimated in-game home team win probability of approximately zero, white an estimated probability of approximately 0.5, and red an estimated probability of approximately one.  Examination of these four figures reveals some common features.  In particular, when the score differential is large, in almost all $(t, \ell)$ cells, all models result in approximately equal estimated probabilities.  Additionally, by the end of the game the estimated win probability goes to either zero or one for all four models.

Estimated probabilities using the traditional MLE and probit models are seen in Figures~\ref{fig:MLE} and~\ref{fig:probit}, respectively.  Both estimates are based on information in a cell, as opposed to a window, and so are more cells with missing data than the Bayesian estimates in Figures~\ref{fig:Bayes} and~\ref{fig:dynamicBayes}.  Additionally, the MLE are less smooth than the other three.  In Figure~\ref{fig:MLE}, up until $t = 1100$, a few windows on the lower bound of the estimates are red, indicating that even though the home team is down early in the game by a very large margin, the probability the home teams wins is approximately one.  These cells illustrate one of the problems in using the MLE.  The estimates produced in these cells are based on a single game.  In February of 2018, Drexel University came back from a 34-point deficit to win over the University of Delaware by a score of 85-83.  This comeback is the ``largest come-from-behind win in the history of Division I basketball'' \citep{espn-2018}.  Delaware led by a score of 53-19 with 2:36 remaining in the first half.  For a many of the cells associated in the first half of this game, both $N_{t,\ell}$ and $n_{t,\ell}$ are one, resulting in a $\bar p_{t, \ell} = 1$.  For this specific game, on these specific cells, the MLE performs perfectly on this training data.  However, as evidenced by other games where the home team fell behind early -- just not as badly as Drexel -- the home team lost with high relative frequency.  In short, cells with low values of $N_{t, \ell}$ result in unreliable estimates because the estimators have large standard errors.  Rare events are not as problematic when using the probit model since the drift $\mu$ and variance $\sigma$ are included in the estimator for each cell.  
The Bayesian estimates perform better estimating probabilities for rare events for two reasons.  The first is seen in a type of interpretation of the scale parameters in the choice of prior.  The first scale parameter can be interpreted as the number of ``pseudo-wins'' in a cell, and the second scale parameter as the number of ``pseudo-losses.''  In this way, the scale parameter choices can be seen as increasing the number of games in a specific window in a way that is intuitive for that window.  Specifically if the home team is ahead, then first scale parameter being large effectively acts to increase the number of wins in that cell.  On the other hand, if the home team is behind, the second scale parameter is large, and thus acting to increase the number of losses.
The second reason rare events do not effect the Bayesian estimates as drastically as the MLE or as the probit model is due to the Bayesian estimates being computed over a small window, as opposed to on a cell.  
Estimates from the Bayes models, shown in Figures~\ref{fig:Bayes} and~\ref{fig:dynamicBayes}, are slower to increase the win probability early in the game compared to the MLE and probit model estimates.  As previously mentioned, because the Bayes estimates are computed using information in a small window around the cell, their overall appearance is smoother.  The most prominent difference in the two Bayesian estimates is for score differentials between $-20$ and 20 points.  In that range, the Bayes estimates using the prior in~\eqref{eq:djprior} tends to result in moderate probabilities (0.4 to 0.6) for a early in the game, and for longer period of elapsed time than the Bayes estimates with the dynamic prior.

\begin{figure}[!ht]
\caption{Adjusted dynamic Bayesian estimates for different values of pre-game win probability.}
  \begin{subfigure}[h]{0.325\textwidth}
    \caption{$\hat p_p = 0.67$.}  
    \includegraphics[width=\textwidth]{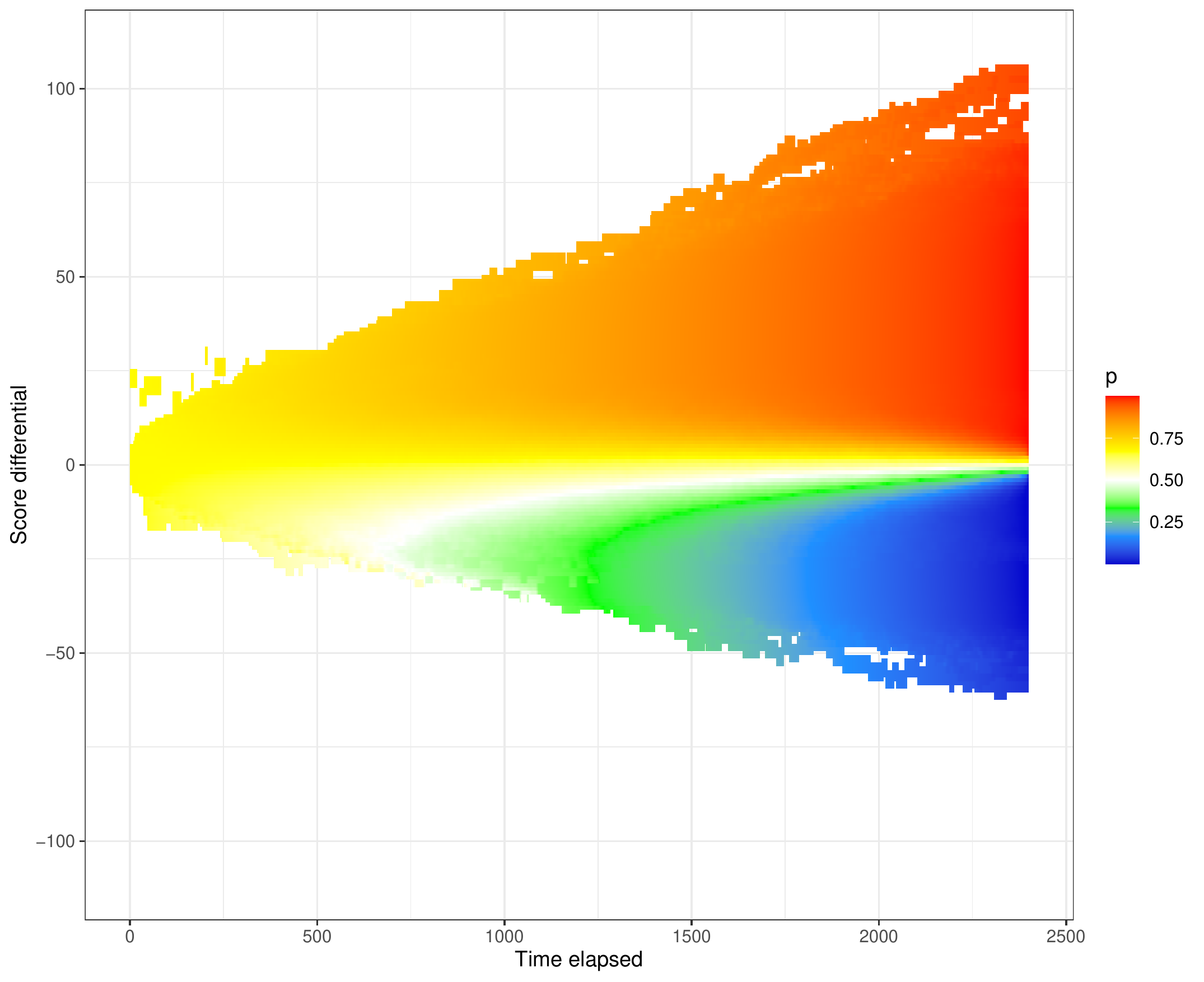}
    \label{fig:aveadjBayes}
  \end{subfigure}
  \hfill
  \begin{subfigure}[h]{0.325\textwidth}
    \caption{$\hat p_p = 0.97$.}
    \includegraphics[width=\textwidth]{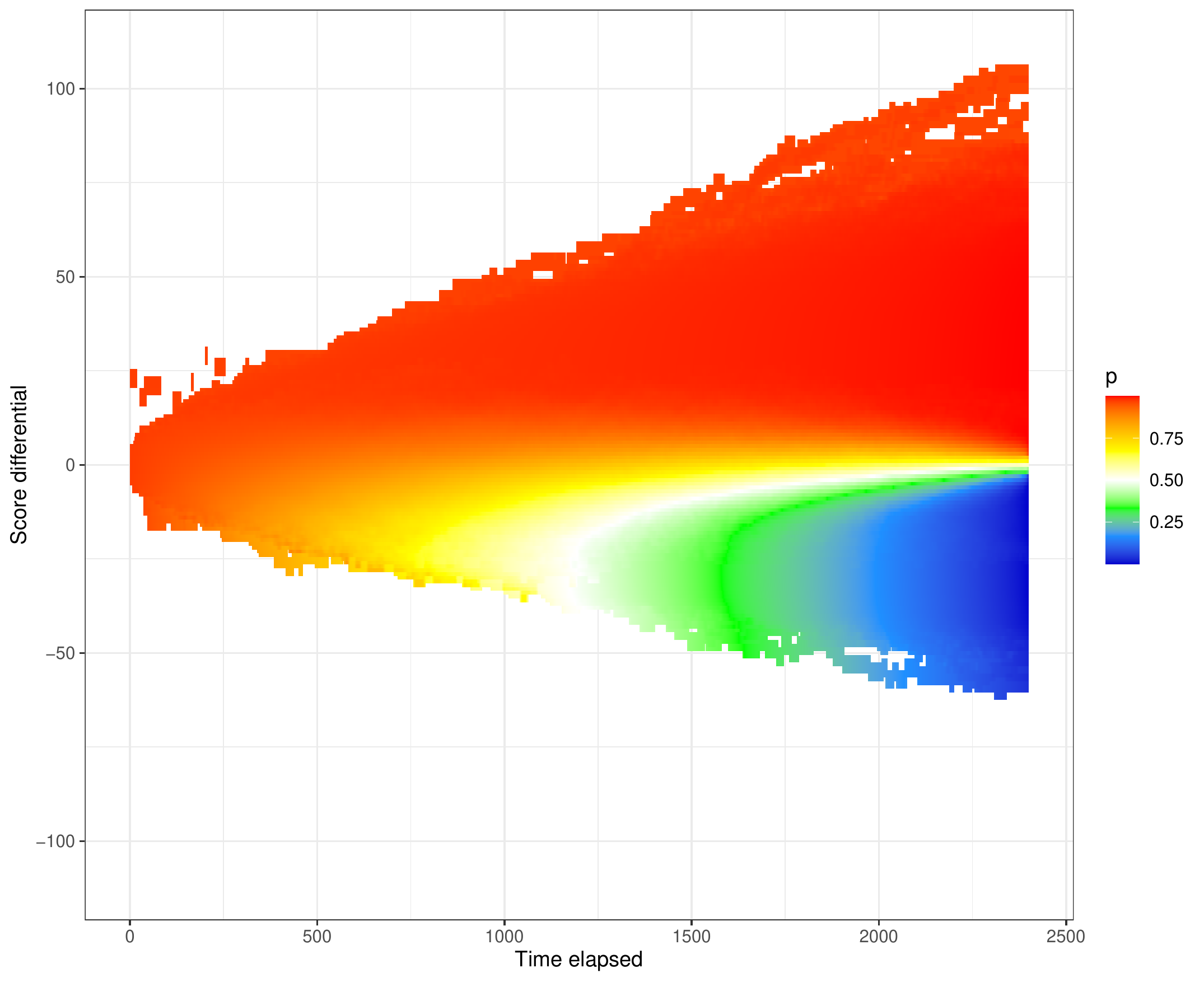}
    \label{fig:homeadjBayes}
  \end{subfigure}
  \hfill
  \begin{subfigure}[h]{0.325\textwidth}
    \caption{$\hat p_p = 0.37$.}
    \includegraphics[width=\textwidth]{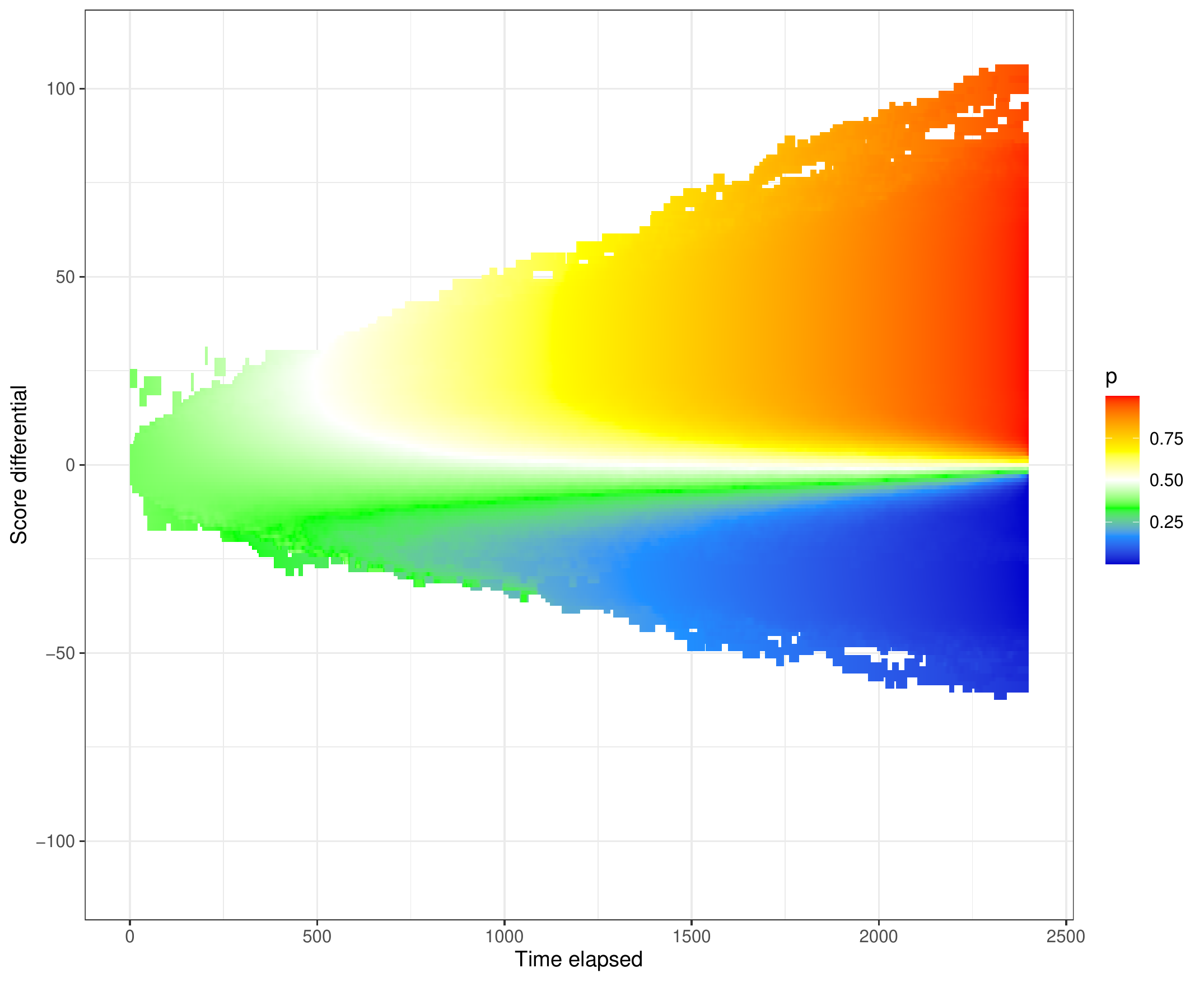}
    \label{fig:roadadjBayes}
  \end{subfigure}
\end{figure}

Figures~\ref{fig:aveadjBayes} through~\ref{fig:roadadjBayes} illustrate the estimated in-game home team win probabilities using the adjusted Bayes estimator with dynamic prior.  Because the estimated probability is affected by the pre-game home team win probability, three values of $\hat p_p$ were selected to illustrate the performance of the newly proposed estimator.  Of the 30,789 games in the data, approximately 67\% were won by the home team, motivating the choice of $\hat p_p = 0.67$.  The other two choices of pre-game home team win probability were found by adding and subtracting 0.3 from 0.67.  For all of the 30,789 games, the adjusted Bayes estimator of equation~\eqref{eq:best} was applied.   For a pre-game win probability of $\hat p_p = 0.67$, the adjusted in-game home team win probabilities are see in Figure~\ref{fig:aveadjBayes}.  Figures~\ref{fig:homeadjBayes} are the estimates for games with a pre-game win probability of 0.97, and Figure~\ref{fig:roadadjBayes} for those with $\hat p_p = 0.37$. A visual comparison of the graphs in Figures~\ref{fig:MLE} through~\ref{fig:Bayes} to Figures~\ref{fig:aveadjBayes} through~\ref{fig:roadadjBayes} leads to the conclusion that the unadjusted estimates perform very differently than the adjusted estimates.  In particular, the adjusted estimates tend to change more slowly than the adjusted estimates.  This is especially true in the first half of the game.  The implications of this slower change is that the adjusted estimates are less likely to produce a high win probability early in the game for large leads.  Later in the game, the adjusted estimates change more rapidly. An animated Shiny app showing the evolution of the adjusted dynamic Bayes estimates as a function of $\hat p_p$ can be found online at {\tt{jasontmaddox.shinyapps.io$\backslash$Win\_Probability}}.

\subsection{Assessing the Models Predictive Performance}
\label{sec:prediction}
The five models fitted in Section~\ref{sec:estimation} were used to predict the outcome of the 10,853 games in 2018-2019 season and the 2019-2020 season, which was cut short due to COVID.  Two of the more common model evaluation methods used in classification of binary prediction models are Brier's score and misclassification rates.  The predictive performances of the MLE, probit model, Bayes with prior in~\eqref{eq:djprior}, Bayes with dynamic prior, and adjusted Bayes model with dynamic prior are compared below using the two measures.
        
Brier's score is analogous to the sum of the squared errors in linear regression and has the advantage of maintaining the continuous information of all estimations.  The square of the difference between the estimated probability is compared to the observed binary outcome.  In the context of in-game home team win probabilities, this observed binary outcome is denoted $y_i$ and is the observed value of $Y_i$ as defined in Section~\ref{sec:models}.  To interpret Brier's score, if $Y_i = 1$ for all $i$, and the predicted probability is also one for every $i$, then Brier's score will be zero, indicating perfect prediction.  On the other hand, if for all $i, Y_i = 0$, and the estimated probability is one, then Brier's score will be one, the worst possible Brier's score.  

To compute Brier's score in this context, let $\tilde p_{t,\ell}$ represent the estimated in-game home team win probability for any one of the five methods.  For each $(t, \ell)$ cell in the test data set consisting of the 10,853 games, let $N^*_{t,\ell}$ represent the number of games in the cell in which the home team led by $\ell$ points at time $t$; that is, $N^*_{t,\ell}$ is the number of games observed in that cell.  Then for non-missing estimated probabilities, Brier's score is
\begin{equation*}
B = \frac{1}{Q}\sum_{t=0}^{2399}\sum_{\ell=-104}^{104}\sum_{j=1}^{N^*_{t,l}} \left(\tilde p_{t,\ell} - y_j\right)^2,
\end{equation*}
where $Q$ is the sum of $N^*_{t,\ell}$ in cells without missing $\tilde p_{t,\ell}$.  Values of $Q$ for the models are given in Table~\ref{tab:Q}.
\begin{table}[!ht]
  \caption{Number of points in computing Brier's score and the misclassification rates for five estimation methods.}
   \begin{tabular}{lccccc} 
         & \multicolumn{5}{c}{Model} \\ \cline{2-6} 
         & MLE & Probit & Bayes & Dynamic Bayes & Adjusted Dynamic Bayes \\ \hline
    $Q$  & 26,034,467 & 26,034,467 & 26,044,933 & 26,047,200 & 26,047,200 \\ 
  \end{tabular}
  \label{tab:Q}
\end{table}

Misclassification rates transform the estimated probabilities into binary indicators. Specifically, a false positive occurs if the estimated in-game home team win probability is greater than 0.5 and the home team loses ($Y_i = 0$).  A false negative occurs when the estimated in-game home team win probability is less than 0.5 and the home team wins ($Y_i = 1$).  According to the same summing operations for calculating Brier's score, the number of false positives $FP$ and false negatives $FN$ are counted, and then the two values are added.  Using this, the misclassification rate is defined by
\begin{equation*}
    MR = \frac{FP + FN}{Q}.
\end{equation*}

Because Brier's score maintains the values of the estimated
probabilities, it conveys more information about the models ability to
predict than the misclassification rate. However, due to the sum of
squares in Brier's score, misclassification rate is more easily
interpreted.  Both of the evaluation metrics for each of the three of
the four models are provided in Table~\ref{tab:prediction}.
\begin{table}[!ht]
    \caption{Evaluation of predictive performances the 2018-2019 and 2019-2020 seasons.}
\begin{tabular}{rcc|cc} 
                                      & \multicolumn{2}{c|}{2018-2019 Season} & \multicolumn{2}{c}{2019-2020 Season} \\ \cline{2-5}
    Model                             & $B$  & $MR$  & $B$ & $MR$  \\ \hline
    MLE                               & 0.1453 & 0.2183 & 0.1397 & 0.2084 \\
    Probit                            & 0.1452 & 0.2180 & 0.1398 & 0.2086 \\
    Bayes                             & 0.1451 & 0.2182 & 0.1396 & 0.2081 \\
    Bayes with dynamic prior          & 0.1450 & 0.2182 & 0.1396 & 0.2081 \\
    Adjusted dynamic Bayes            & 0.1284 & 0.1870 & 0.1261 & 0.1827 \\ 
\end{tabular}
    \label{tab:prediction}
\end{table}
With respect to these metrics, the MLE, probit model, Bayes model, and Bayes model with dynamic prior appear to perform similarly.  However, with values of $Q$ exceeding 26,000,000, these slight differences may be indicative of a true difference in performance. In contrast, the adjusted Bayesian model with dynamic prior clearly out performs the other methods, having both lower Brier's score, and misclassification rate.

\section{Application of Model to a Specific Game}
\label{sec:application}
\begin{figure}[!ht]
\caption{Win probability graph for 2016 NCAA Championship Game.}
\includegraphics[width=0.8\textwidth]{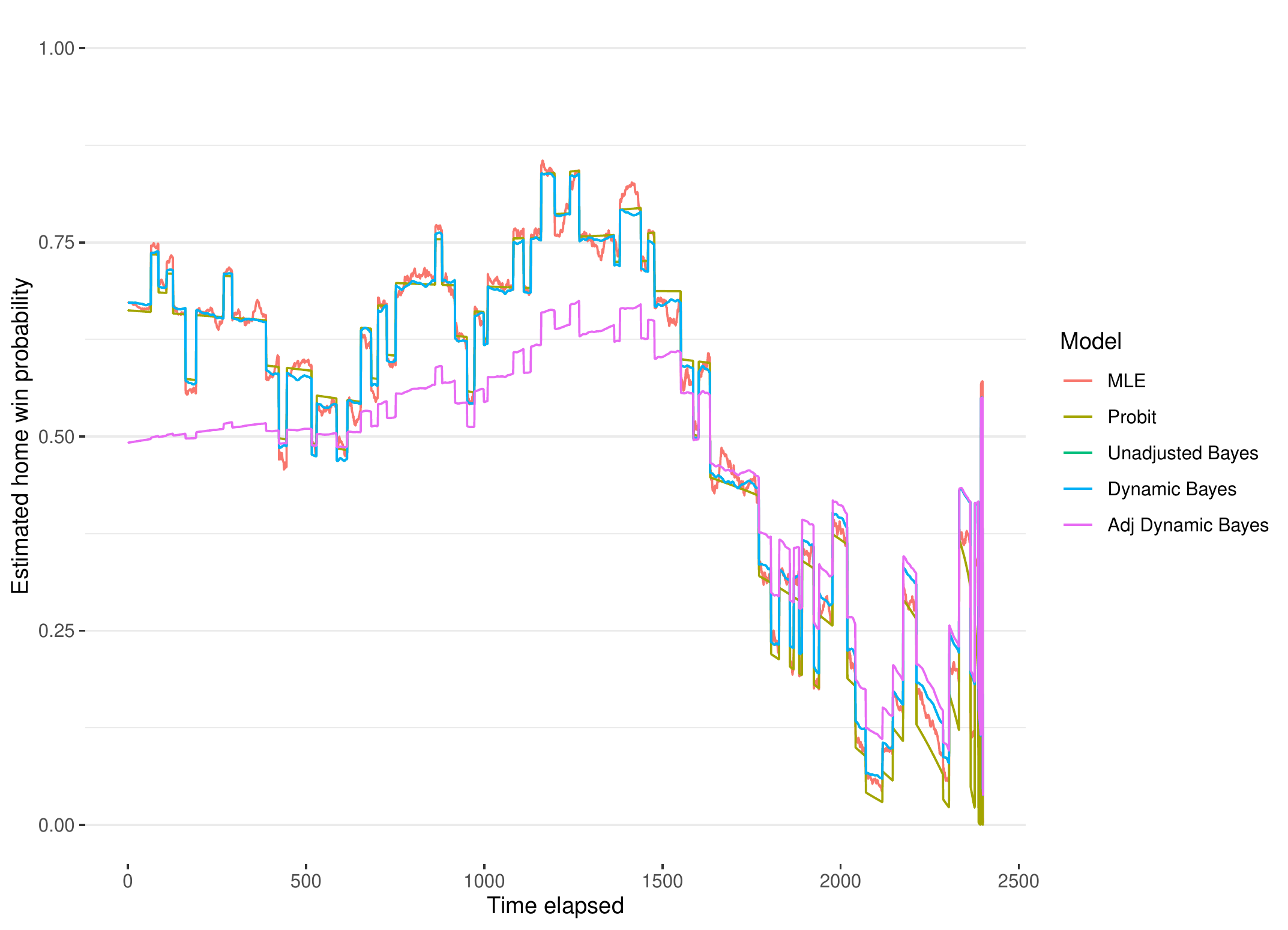}
\label{fig:application}
\end{figure}
To illustrate utility for a single game, the five estimation techniques were applied to the 2016 NCAA Division 1 Championship Game between the University of North Carolina (UNC) and Villanova University.  The traces of the estimated in-game home team win probabilities for all five methods are seen in Figure~\ref{fig:application}. The game was played at NRG Stadium in Houston, Texas, a neutral site.  UNC was listed as the home team, as one of the four \#1 seeds in the Tournament that year.  Villanova was a \#2 seed, but was favored slightly over UNC by {\tt teamrankings.com}.  The estimated pre-game win probability for UNC to win the game was $\hat p_p = 0.49$.  The first three points of the game were scored one minute and five seconds into the game by UNC's Joel Berry II.  Nineteen seconds later, Kris Jenkins of Villanova made a lay-up, and the score was 2-3.  The game stayed close, and at the end of the first half (1,200 seconds into the game), UNC led Villanova 39-34.  An investigation of the five estimators shows that the MLE, probit model, Bayes, and unadjusted dynamic Bayes all quickly react to the early scores, bumping up the UNC win probability to as high as 0.75. At the half, when UNC's lead was five points, the four unadjusted methods predict UNC will win with a probability of around 0.85.  On the other hand, the adjusted Bayesian estimate does not react as drastically as the other four in the early part of the game, because the pre-game win probability is influencing the returned values.  

Five minutes and 52 seconds (1,552 seconds) into the second half, Villanova had tied the game.  At that time, the estimated probability of a UNC win drops to around 0.5 for all five methods.  As the game continued, and the clock wound down, the five estimates increase and decrease according to the score differential.  In the last few minutes of the game, the change in the five estimates is more unified in response to the change in score differential than in the first half of the game.  At around time 2,370 seconds elapsed, Villanova had an eight point lead, and all five models give UNC a very small chance of winning.  UNC pulled within one point with one minute remaining.  With six seconds remaining in the game, Marcus Page of UNC hit a 3-point jump shot to tie the game at 74-74, and all five estimators jump to a home team win probability of greater than 0.5.  With less than one second remaining, Kris Jenkins shot a 3-pointer which went through the net after time expired. Villanova won the championship by a score of 77-74, and all five estimators drop to zero, showing how Villanova taking the lead with no time remaining drastically affects the home team's (UNC) probability of winning.  

\section{Conclusion}
\label{sec:conclusion}
For NCAA basketball games, two new methods are proposed for estimating or predicting in-game home team win probabilities.  The first newly proposed method is a Bayesian estimator with a prior distribution that changes as a function of lead differential and time elapsed, which was called the Bayesian estimator with a dynamic prior.  The second method adds to the original estimate a time-weighted adjustment based on pre-game win probability computed from daily ratings.  In this paper, the adjustment was applied only to the Bayesian estimate with dynamic prior. It is reasonable to conclude the adjustment would improve the performance of the other estimators, just as it did the dynamic Bayesian estimator.  A comparison of the methods for the purpose of estimation shows that the two proposed estimates out-performs the estimates from the three standard methods.  For prediction, the adjusted dynamic Bayesian method out performs the other, based on a comparison of both Brier's score and misclassification rates.  There are a number of additional problems to be investigated.  First is the determination of an appropriate method for computing pre-game win probabilities so the same adjustments might be applied to probability estimates for NBA games.  Also of interest is the effect on the estimators for different window choices.  Another consideration is that the adjustment resulted in a substantial improvement on the prediction of the adjusted dynamic Bayesian estimator.  A different weighting function may further improve the predictions.  Finally, the development of these types of models for other sports also presents unique challenges that are worth investigating.
\bibliographystyle{apalike}
\bibliography{refs}
\end{document}